\documentclass{article}

\usepackage{arxiv}

\usepackage[utf8]{inputenc} 
\usepackage[T1]{fontenc}    
\usepackage{hyperref}       
\usepackage{url}            
\usepackage{booktabs}       
\usepackage{amsmath}       
\usepackage{nicefrac}       
\usepackage{microtype}      
\usepackage{lipsum}
\usepackage{multirow,bm}
\usepackage{graphicx}
\usepackage{geometry}
\usepackage{array}
\usepackage{algorithm,algorithmic}
\usepackage{xcolor}
\graphicspath{ {./images/} }

\usepackage{titlesec}
\titleformat{\paragraph}
{\normalfont\normalsize\bfseries}{\theparagraph}{1em}{}
\titlespacing*{\paragraph}
{0pt}{3.25ex plus 1ex minus .2ex}{1.5ex plus .2ex}

\title{Calibrated and Enhanced NRLMSIS 2.0 Model with Uncertainty Quantification}

\author{
 Richard J. Licata \\
  Dept. of Mechanical and Aerospace Engineering \\
  West Virginia University\\
  Morgantown, WV 26505 \\
  \texttt{rjlicata@mix.wvu.edu} \\
   \And
 Piyush M. Mehta \\
  Dept. of Mechanical and Aerospace Engineering \\
  West Virginia University \\
  Morgantown, WV \\
     \And
 Daniel R. Weimer\;\;\;\;\; \\
  Center for Space Science and Eng. Research\;\;\;\;\; \\
  Virginia Tech\;\;\;\;\; \\
  Blacksburg, VA\;\;\;\;\; \\
     \And
   W. Kent Tobiska\;\;\;\;\;\; \\
  Space Environments Technologies\;\;\;\;\;\; \\
  Pacific Palisades, CA\;\;\;\;\;\; \\
       \And
 Jean Yoshii \\
  Space Environments Technologies \\
  Pacific Palisades, CA \\
}

\begin{document}
\maketitle
\begin{abstract}

The Mass Spectrometer and Incoherent Scatter radar (MSIS) model family has been developed and improved since the early 1970's. The most recent version of MSIS is the Naval Research Laboratory (NRL) MSIS 2.0 empirical atmospheric model. NRLMSIS 2.0 provides species density, mass density, and temperature estimates as function of location and space weather conditions. MSIS models have long been a popular choice of atmosphere model in the research and operations community alike, but -- like many models -- does not provide uncertainty estimates. In this work, we develop an  exospheric temperature model based in machine learning (ML) that can be used with NRLMSIS 2.0 to calibrate it relative to high-fidelity satellite density estimates. Instead of providing point estimates, our model (called MSIS-UQ) outputs a distribution which is assessed using a metric called the calibration error score. We show that MSIS-UQ debiases NRLMSIS 2.0 resulting in reduced differences between model and satellite density of 25\% and is 11\% closer to satellite density than the Space Force's High Accuracy Satellite Drag Model. We also show the model's uncertainty estimation capabilities by generating altitude profiles for species density, mass density, and temperature. This explicitly demonstrates how exospheric temperature probabilities affect density and temperature profiles within NRLMSIS 2.0. Another study displays improved post-storm overcooling capabilities relative to NRLMSIS 2.0 alone, enhancing the phenomena that it can capture.

\end{abstract}

\section{Introduction}\label{sec:intro}

The thermosphere is the neutral portion of the upper atmosphere where many satellites, debris, and space assets reside. It is primarily heated and influenced by solar extreme ultraviolet (EUV) and far ultraviolet (FUV) emissions which vary with the solar cycle \cite{Emmert07}. Geomagnetic storms -- often caused by coronal mass ejections (CMEs) and coronal holes -- can cause sudden increases in mass density with little warning. In early 2022, SpaceX had 38 satellites fail to reach their desired orbits due to a density response to a CME which resulted in a minor geomagnetic storm \cite{spacex}. While the occurrence could have been avoided with adequate neutral density forecasts, current models carry high errors and uncertainty during geomagnetic events \cite{current_state}.

Although geomagnetic storms receive considerable attention due to their immediate consequences, density model performance also varies as a function of solar activity. Bowman et al. \cite{JB2008} compared density ratios and error standard deviations of thermosphere models with respect to the High Accuracy Satellite Drag Model (HASDM) which is the operational model used by the United States Space Force \cite{HASDM}. Their results showed considerable variability as a function of solar activity with strong underestimation (density ratio of 0.7--0.8) during solar minimum. Furthermore, the error standard deviation of these models can be up to 45\% during solar minimum and no less than 10\% during solar maximum. This points to the need for innovative solutions to provide overall improvements to our thermospheric mass density models.

One such solution is to develop a correction-based model using satellite density estimates \cite{intercal,ruan,Choury,EXTEMPLAR,extemplar_ml}. Over the last few decades, there have been a growing number of satellites with high-fidelity onboard accelerometers such as CHAllenging Minisatellite Payload (CHAMP) \cite{CHAMP}, Gravity Recovery and Climate Experiment (GRACE) \cite{grace}, and Swarm \cite{Swarm}. Some researchers have taken the accelerometer data from the satellites, removed other acceleration sources (e.g. solar radiation pressure), and isolated drag acceleration therefore estimating density \cite{CHden1,CHden2,Sutton,Doorn,EOF3,CHAGRA,Swarm_dens}. These density sources have a high temporal cadence and, when combined, provide coverage over a wide array of locations and space weather conditions.

In this work, we develop an exospheric temperature model based on satellite estimates using machine learning (ML) to feed into NRLMSIS 2.0 \cite{MSIS2}. This model (called MSIS-UQ) provides a distribution in its exospheric temperature predictions therefore incorporating UQ capabilities to NRLMSIS 2.0. MSIS-UQ differs from EXospheric TEMperatures on a PoLyhedrAl gRid - ML (EXTEMPLAR-ML) by: 1) using true locations (no grid) for training, 2) using the newest MSIS model, and 3) providing uncertainty estimates. The manuscript is organized as follows. We describe the data, model development, and validation approaches. We then show results for overall model performance and a demonstration of its uncertainty capabilities. Last, we evaluate MSIS-UQ during a storm and perform a study to test its response to geomagnetic activity.

\section{Data and Methods}
\subsection{Exospheric Temperature Estimates}\label{sec:temp_est}

Exospheric temperatures were obtained through a binary search, deriving the exospheric temperature required in NRLMSIS 2.0 to match satellite density estimates \cite{intercal,comparison}. The density estimates originate from the following sources -- CHAMP 2001: Doornbos \cite{Doorn}, CHAMP 2002-2010: Mehta et al. \cite{CHAGRA}, GRACE 2002-2009: Mehta et al. \cite{CHAGRA}, GRACE 2010: Sutton \cite{Sutton}, Swarm A 2013-2018: Astafyeva et al. \cite{Swarm_dens}, and Swarm B 2012-2016: Astafyeva et al. \cite{Swarm_dens}. Note that only GRACE-A measurements are used due to the similarity of the GRACE-A and GRACE-B orbits. The CHAMP and GRACE density estimates originate from accelerometer measurements and span an altitude range of 300--535 km while the Swarm A and B density estimates are obtained from GPS data and span an altitude range of 437--545 km. The cadence of the satellite estimates are 10 s, 5 s, 30 s, and 30 s for CHAMP, GRACE, Swarm A, and Swarm B, respectively. There are over 82 million samples total for model development and evaluation.

\subsection{Model Drivers}\label{sec:drivers}

To run NRLMSIS 2.0, the following space weather inputs are required: \textit{F\textsubscript{10.7}}, \textit{F\textsubscript{81c}}, and \textit{Ap}. \textit{F\textsubscript{10.7}} is a measure of the 10.7 cm solar radio flux and acts as a good proxy for solar EUV emissions while \textit{F\textsubscript{81c}} is merely an 81-day centered average of the proxy. The \textit{Ap} index is a measure of daily global geomagnetic activity. NRLMSIS 2.0 also has a storm-time flag that allows additional 3-hourly \textit{ap} values: \textit{ap}, \textit{ap\textsubscript{3}}, \textit{ap\textsubscript{6}}, \textit{ap\textsubscript{9}}, \textit{ap\textsubscript{12-33}}, and \textit{ap\textsubscript{33-57}}. The subscripts indicate the number of hours prior to epoch that \textit{ap} value represents; when there are two numbers (e.g. 33-57), the 3-hourly index values are averaged over this many hours prior to epoch. NRLMSIS also has temporal inputs (e.g. time of day, day of year).

To develop a machine-learned exospheric temperature model with a high temporal cadence, we expand the input set. To account for solar activity, the model receives \textit{F\textsubscript{10.7}}, \textit{S\textsubscript{10.7}}, \textit{M\textsubscript{10.7}}, and \textit{Y\textsubscript{10.7}}, all accounting for different forms of solar emissions that affect different regions of the thermosphere \cite{dev_inds,JB2008}. The ML model also uses inputs from EXTEMPLAR, particularly \textit{S\textsubscript{N}}, \textit{S\textsubscript{S}}, $\Delta T$. The two \textit{S} inputs are Poynting flux totals in the Northern and Southern hemispheres \cite{W05a,W05b}. $\Delta T$ is a parameter derived by Weimer et al. \cite{EXTEMPLAR,comparison} and represents exospheric temperature perturbations; it is a function of Poynting flux and simulated nitric oxide emissions.

For further representation of geomagnetic activity, a time history of \textit{SYM-H} is used, similar to the use of storm-time \textit{ap} indices for NRLMSIS 2.0. \textit{SYM-H} represents disturbances to the geomagnetic field and has similar characteristics to \textit{Dst}. A benefit over \textit{Dst} is its 1-min cadence \cite{symh}. The \textit{SYM-H} inputs are as follows:  \textit{SYM-H}, \textit{SYM-H\textsubscript{0-3}}, \textit{SYM-H\textsubscript{3-6}}, \textit{SYM-H\textsubscript{6-9}}, \textit{SYM-H\textsubscript{9-12}}, \textit{SYM-H\textsubscript{12-33}}, \textit{SYM-H\textsubscript{33-57}}. Due to the model's use of in-situ satellite estimates, it takes location as an input. The local solar time (LST) is transformed using sine and cosine functions to make it continuous about midnight,
\begin{equation} \label{eqLST}
\begin{split}
LST_1=sin\left(\frac{2\pi LST}{24}\right)
\;\;\;\;
LST_2=cos\left(\frac{2\pi LST}{24}\right)
\end{split}
\end{equation}
The model also uses satellite latitude (\textit{LAT}). To account for temporal dependencies, day of year (\textit{t\textsubscript{1}} and \textit{t\textsubscript{2}}) and universal time (UT, \textit{t\textsubscript{3}} and \textit{t\textsubscript{4}}) are transformed through similar functions,
\begin{equation} \label{eqT}
\begin{split}
t_1=sin\left(\frac{2\pi doy}{365.25}\right)
\;\;\;\;
t_2=cos\left(\frac{2\pi doy}{365.25}\right)
\;\;\;\;
t_3=sin\left(\frac{2\pi UT}{24}\right)
\;\;\;\;
t_4=cos\left(\frac{2\pi UT}{24}\right)
\end{split}
\end{equation}
\subsection{Data Preparation}\label{sec:prep}

The ML model drivers are the 21 space weather, spatial, and temporal inputs mentioned in the previous section. Each sample has an exospheric temperature estimate which will serve as the target. To reduce the variance of the output ($T_{\infty}$), we use a logarithmic transformation making the output $log_{10}T_{\infty}$. The final step to create ML-ready data is to perform standard normalization. This will center the data and provide a unit standard deviation for each input and output with respect to the statistics of the training data,
\begin{equation} \label{eqnorm}
\begin{split}
\widetilde{x} = \frac{x-\mu}{\sigma}
\end{split}
\end{equation}
where \textit{x} represents each input and the output, $\mu$ is the mean of that quantity over the training set, and $\sigma$ is the standard deviation of that quantity over the training set. $\mu$ and $\sigma$ must be saved for downstream use of the model. 

The 81 million samples are split into training, validation, and tests sets to achieve an 80\%--10\%--10\% distribution. Licata et al. \cite{extemplar_ml} split a similar dataset to have long segments -- on the order of months or years. However, the resulting model was not well-generalized across the sets. When comparing EXTEMPLAR and a previous version of MSIS over the same three time periods, the error statistics also varied. Therefore, we use a different approach to data splitting. Starting with the first sample, eight weeks are used for training, the following week is used for validation, and the subsequent week is used for testing. This pattern is repeated until the end of the dataset. This approach forces similar solar cycle and spatial coverage across the three sets. Concurrently, there is a significant number of samples within each segment providing temporally disjoint segments throughout the 17 year time-span of the dataset. Since each satellite has a different cadence and there are different numbers of satellites providing measurements at a given time, the number of samples in each segment varies. In the training, validation, and test sets, the number of samples varies from 22,454--1,450,380, 12,990--181,423, and 12,888--181,422, respectively. This means that there are between 25,878 and 362,845 samples separating training segments.

\subsection{Model Development}\label{sec:MLdev}

A key consideration in ML model development is the choice for the loss (or objective) function. This will be the quantity that the algorithm will try to minimize or maximize during the training phase. For this work, we use the negative logarithm of predictive density (NLPD) loss function due to previous ML modeling results \cite{HASDM_ML,uq_tech}. NLPD is given as
\begin{equation} \label{eqNLPD}
\begin{split}
NLPD(y,\mu,\sigma) = \frac{\left(y-\mu\right)^2}{2\sigma^2}+\frac{log\left(\sigma^2\right)}{2} + \frac{log(2\pi)}{2}
\end{split}
\end{equation}
where $y$, $\mu$, and $\sigma$ are the ground truth, mean prediction, and predicted standard deviation, respectively. This loss functions provides the capability for uncertainty prediction since the model can now make $\sigma$ predictions and have them incorporated in the loss function without labels. The ML model will directly predict $\mu$ and $\sigma$ for $log_{10}T_{\infty}$ meaning it has two outputs. The output node for $\mu$ uses a linear activation function since the normalized values can take on any value (unbounded). The output node for $\sigma$ must be positive, monotonically increasing, and have no upper-bound. Therefore, we proceed with the "softplus" activation function: $f(x) = ln(1+e^x)$.

With the inputs/output(s) selected, normalization complete, loss function chosen, and output layer determined, the only remaining task for model development is to select an architecture and train a model. To accomplish this, we defined a scheme and hyperparameter space using Keras Tuner \cite{KT} as with previous work \cite{extemplar_ml,HASDM_ML,uq_tech}. Table \ref{t:KT} shows the parameters used and ranges provided to the tuner. Each hidden layer can have its own unique number of neurons, activation function, and dropout rate. Note: "trials" refers to number of model architectures, initial points is the number of randomly selected architectures prior to Bayesian optimization, and repeats refers to re-initializing weights and retraining.

\begin{table}[htb!]
	\fontsize{10}{10}\selectfont
    \caption{Hyperparameter tuner parameters (left) and search space (right).}
   \label{t:KT}
        \centering 
   \begin{tabular}{ c | c | c | c} 
      \hline 
            \textbf{Tuner Option} & \textbf{Choice} & \textbf{Parameter} & \textbf{Values/Range} \\ \hline 
            \multirow{2}{*}{\textit{Scheme}} & Bayesian & \textit{Number of} & \multirow{2}{*}{1--10}\\ 
            & Optimization & \textit{Hidden Layers} & \\ \hline 
            \multirow{2}{*}{\textit{Total Trials}} & \multirow{2}{*}{150} & \multirow{2}{*}{\textit{Neurons}} & min = 16, max = 1024,\\
            & & & step = 4 \\\hline 
            \multirow{2}{*}{\textit{Initial Points}} & \multirow{2}{*}{50} & \multirow{2}{*}{\textit{Activations}} & relu, softplus, tanh, sigmoid, \\ 
            & & & softsign, selu, elu\\ \hline
            \multirow{2}{*}{\textit{Repeats per Trial}} & \multirow{2}{*}{3} & \multirow{2}{*}{\textit{Dropout}} & min = 0.10, max = 0.60,\\ 
            & & & step = 0.01 \\ \hline
            \textit{Minimization} & \multirow{2}{*}{Validation Loss} & \multirow{2}{*}{\textit{Optimizer}} & RMSprop, Adam, Adadelta,\\
            \textit{Parameter} & & & Nadam \\
      \hline
   \end{tabular}
\end{table}
Since there are over 65 million training samples in the dataset, we only provide the tuner with a subset of this data. The tuner uses 1 million randomly selected samples from the training set and 200,000 randomly selected samples from the validation set. Each model trained by the tuner will run for 50 training iterations (or epochs) with a batch size of 4,096. Upon completion, the 10 best models are saved based on the validation loss values. All 10 models are evaluated (see Section \ref{sec:analysis}), and the best performing one is used as a base architecture for full training. The best architecture from the tuner is displayed in Table \ref{t:arch}.

\begin{table}[htb!]
	\fontsize{10}{10}\selectfont
    \caption{Model architecture for the best model from the tuner. There are 21 inputs for Layer 1, and it uses the NAdam optimizer \url{https://www.tensorflow.org/api_docs/python/tf/keras/optimizers/Nadam}.}
   \label{t:arch}
        \centering 
   \begin{tabular}{ c | c | c | c } 
      \hline 
             & \textbf{Neurons} & \textbf{Activation} & \textbf{Dropout Rate}\\ \hline 
            \textbf{Layer 1} & 648 & relu & 0.13\\ \hline 
            \textbf{Layer 2} & 176 & tanh & 0.21\\ \hline 
            \textbf{$\mathbf{\mu}$\; Output} & 1 & linear & 0.00 \\ \hline
            \textbf{$\mathbf{\sigma}$\; Output} & 1 & softplus & 0.00 \\
      \hline
   \end{tabular}
\end{table}
\subsection{Model Analysis}\label{sec:analysis}

When comparing model to satellite densities, we use the mean absolute error (MAE) metric in percentage form. This metric is chosen due to its intuitive nature. To assess the quality of the ML uncertainty estimates, we use a calibration error score (CES) \cite{HASDM_ML,uq_tech}. This metric assesses how close the uncertainty estimates are to input prediction intervals (e.g. how close a 95\% prediction interval is to containing 95\% of true samples). The equations for MAE and CES are provided in Equations \ref{eqmae} and \ref{eqces}, respectively.
\begin{subequations}
\begin{equation} \label{eqmae}
\begin{split}
\textrm{MAE} = 100\% \; \frac{\left| \rho_{satellite} - \rho_{model} \right|}{\rho_{satellite}}
\end{split}
\end{equation}
\begin{equation} \label{eqces}
\begin{split}
\textrm{CES} = \frac{100\%}{m \cdot n_{out}}\sum^{n_{out}}_{i=1} \sum^m_{j=1} \Big|p(z_{i,j})-p(\hat{z}_{i,j})\Big|
\end{split}
\end{equation}
\end{subequations}
In Equation \ref{eqmae}, $\rho_{satellite}$ is the estimate of density from a given satellite measurement (e.g. from CHAMP, GRACE, etc.). $\rho_{model}$ is the density from the model being evaluated. In Equation \ref{eqces}, $m$ is the number of prediction intervals being tested, $n_{out}$ is the number of model outputs (for the ML model $n_{out}$ = 1), $p(z_{i,j})$ is a given prediction interval, and $p(\hat{z}_{i,j})$ is the percentage of samples captured by the model's uncertainty estimates with the given prediction interval. In this work, the prediction intervals span from 5\% to 99\% in increments of 5\%.

\subsubsection{Comparison with NRLMSIS 2.0 and HASDM}

To assess the validity of the model in terms of mean density prediction, its error with respect to the satellite estimates are compared to those of NRLMSIS 2.0 and HASDM. To get NRLMSIS 2.0 errors, the model is evaluated at all locations and times of the satellite measurements. For HASDM, the 3-dimension density grids from the SET HASDM density database are interpolated in log-scale to the satellite locations and times \cite{HASDM_SET_Data}. We then break up the errors into the three sets used for ML model development (training, validation, and test). We do this to simultaneously test the generalization of our model while ensuring differences in performance across the sets is also seen with the other models. In addition to the error assessment, we also compute the CES for MSIS-UQ across the three sets (in terms of density). For information on the conversion from ML predicted exospheric temperature to NRLMSIS 2.0 adjusted density, see Weimer et al. \cite{EXTEMPLAR,comparison}.

\subsubsection{Uncertainty Demonstration}\label{sec:UQmethod}

The reliability of the MSIS-UQ uncertainty estimates is demonstrated in Section \ref{sec:results}, but we further establish the capabilities in Section \ref{sec:alt_UQ}. The ML model directly predicts the uncertainty into the exospheric temperature which is then incorporated into NRLMSIS 2.0. The probabilistic $T_\infty$ values result in probabilistic local temperatures and species densities. We consider a given epoch (May 13, 2007 at 21:42.50 UT) where CHAMP and GRACE are at very different locations; CHAMP is near the equator on the night-side while GRACE is at high latitude on the day-side. NRLMSIS 2.0 is provided probabilistic $T_\infty$ values from the MSIS-UQ distribution at each location, and we consider the temperature, species densities, and mass density between 100 and 800 km altitude. The distributions are shown as a function of altitude, and the satellite estimates are provided for reference.

\subsubsection{Post-storm Cooling Capabilities}

With overall performance investigated, we want to consider the 2003 Halloween storm along the CHAMP orbit. We interpolate global density grids for TIE-GCM \cite{TIEGCM}, JB2008 \cite{JB2008}, and HASDM \cite{HASDM}. In addition, we evaluate NRLMSIS 2.0, EXTEMPLAR, and MSIS-UQ directly at the CHAMP locations. The orbit-averaged densities are computed for all models (and CHAMP) for display and comparison. This can provide insight into storm responses and post-storm density depletion characteristics. In previous work, we investigated density ratios for NRLMSIS 2.0 and three ML models with time-lagged geomagnetic activity to examine post-storm density characteristics expected within the models \cite{scienceML}. While two of the models exhibited significant density depletion, NRLMSIS 2.0 never showed evidence of thermospheric overcooling. We perform the same study -- outlined below -- to test if the exospheric temperature corrections can enforce the behavior present in the satellite datasets.

To start, all non-geomagnetic model drivers are kept to constant values. We set the solar indices to 120 solar flux units, and the time for the study is 00:00 UT during the fall equinox. Each of the time-history geomagnetic drivers will be increased individually while all others are kept at a constant value: \textit{ap} = 56, \textit{SYM-H} = -50 \textit{nT}. Since the ML model uses Poynting Flux totals and $\Delta T$ at epoch, they are kept constant at 200 GW and 120 K, respectively. This study is also conducted at four discrete locations: night equator (2 hrs \textit{LST}, 0$^\circ$ \textit{LAT}), day equator (14 hrs \textit{LST}, 0$^\circ$ \textit{LAT}), night pole (2 hrs \textit{LST}, 80$^\circ$ \textit{LAT}), and day pole (14 hrs \textit{LST}, 80$^\circ$ \textit{LAT}). The density ratios are reported with respect to the given geomagnetic driver being set to 0. 

However, achieving this requires an additional consideration relative to the work of Licata et al. \cite{scienceML}. MSIS-UQ uses \textit{SYM-H} while NRLMSIS 2.0 uses \textit{ap} for time-series geomagnetic drivers. To account for this distinction, we first fit a line between all \textit{SYM-H} and \textit{ap} values within our dataset. Using this, we find the \textit{SYM-H} value associated with the \textit{ap} value that must be used to get density from NRLMSIS 2.0. Therefore, density ratios for MSIS-UQ use this simultaneous \textit{SYM-H} and \textit{ap} variation as opposed to only using \textit{ap} variations with NRLMSIS 2.0 alone.

\section{Results} \label{sec:results}

\begin{figure}[htb!]
	\centering
	\small
	\includegraphics[width=\textwidth]{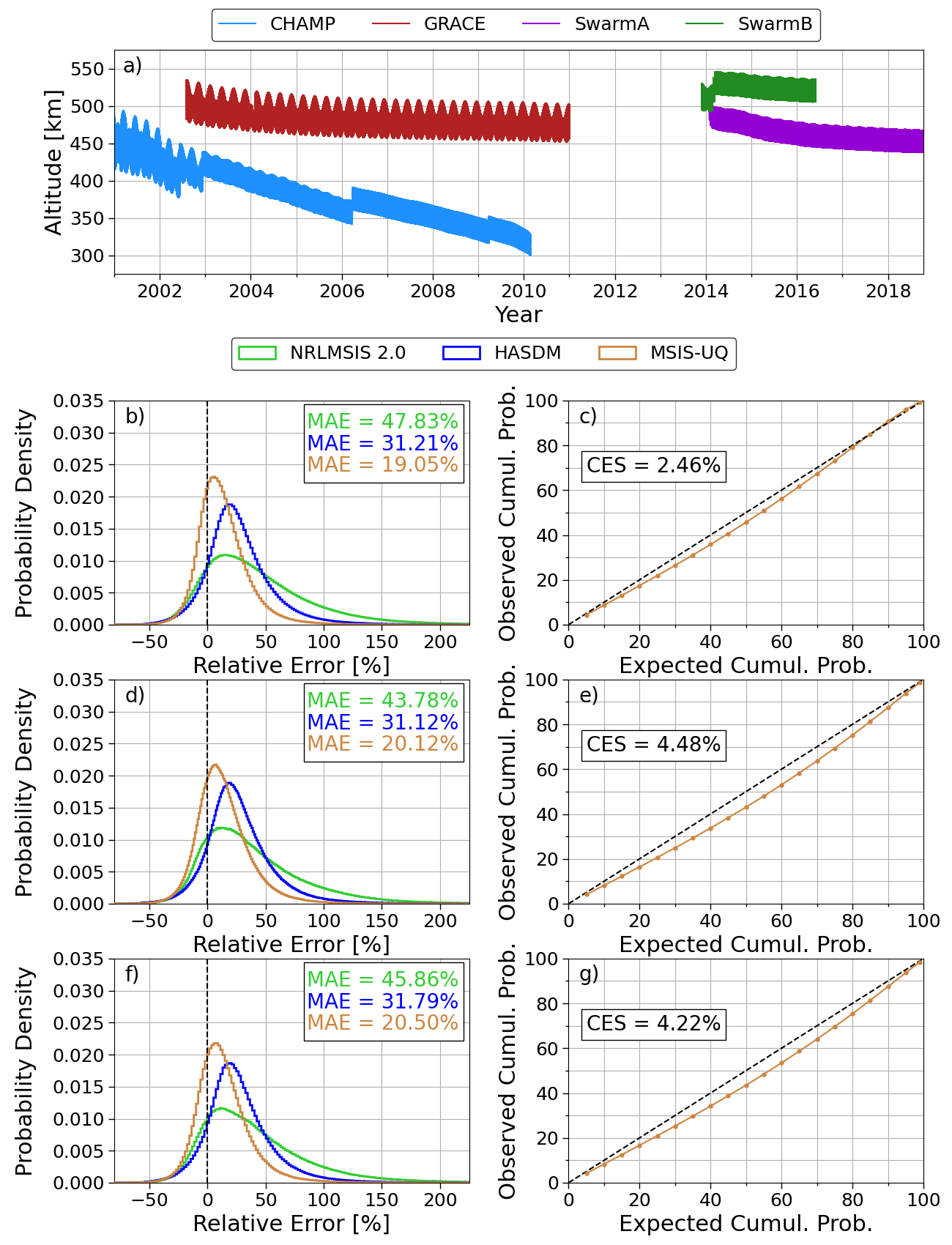}
    \caption{Altitudes of the satellites used for temperature and density estimates (a), relative error histograms on training set (b), MSIS-UQ training set calibration curve (c). relative error histograms on validation set (d), MSIS-UQ validation set calibration curve (e), relative error histograms on test set (f), MSIS-UQ test set calibration curve (g).}
	\label{f:TVT}
\end{figure}

Figure \ref{f:TVT} shows the relative error distributions and mean absolute error for NRLMSIS 2.0, HASDM, and MSIS-UQ with respect to density estimates from CHAMP, GRACE, Swarm A, and Swarm B. The calibration curve for MSIS-UQ is also displayed alongside the calibration error score. This is separated by samples in the MSIS-UQ training, validation, and test sets. Similar figures are provided for each individual satellite in SM1. Note: these statistics are relative errors for MSIS-UQ since it was developed on this dataset, but they could be interpreted as relative differences for NRLMSIS 2.0 and HASDM.

Panel (a) shows the altitudes for each satellite used in this analysis showing over a 200 km span over 15 years of measurements. The left-most panels (b), (d), and (f) indicate that MSIS-UQ provides much more accurate density predictions than both NRLMSIS 2.0 alone and HASDM. All three models have a tendency to overpredict density although MSIS-UQ has the smallest bias. The MAE values highlight the $\sim$25\% error reduction from NRLMSIS 2.0 and the $\sim$11\% error reduction from HASDM. Across the three sets, MSIS-UQ is well-generalized with density prediction errors ranging <1.5\%. With respect to its uncertainty estimates (panels (c), (e), and (g)), MSIS-UQ has a CES < 5\% across the three sets. It has a tendency to underestimate in the middle prediction intervals (between 20\% and 80\%) but is well-calibrated at prediction intervals > 90\%.

\subsection{Uncertainties as a Function of Altitude}\label{sec:alt_UQ}

\begin{figure}[htb!]
	\centering
	\small
	\includegraphics[width=\textwidth]{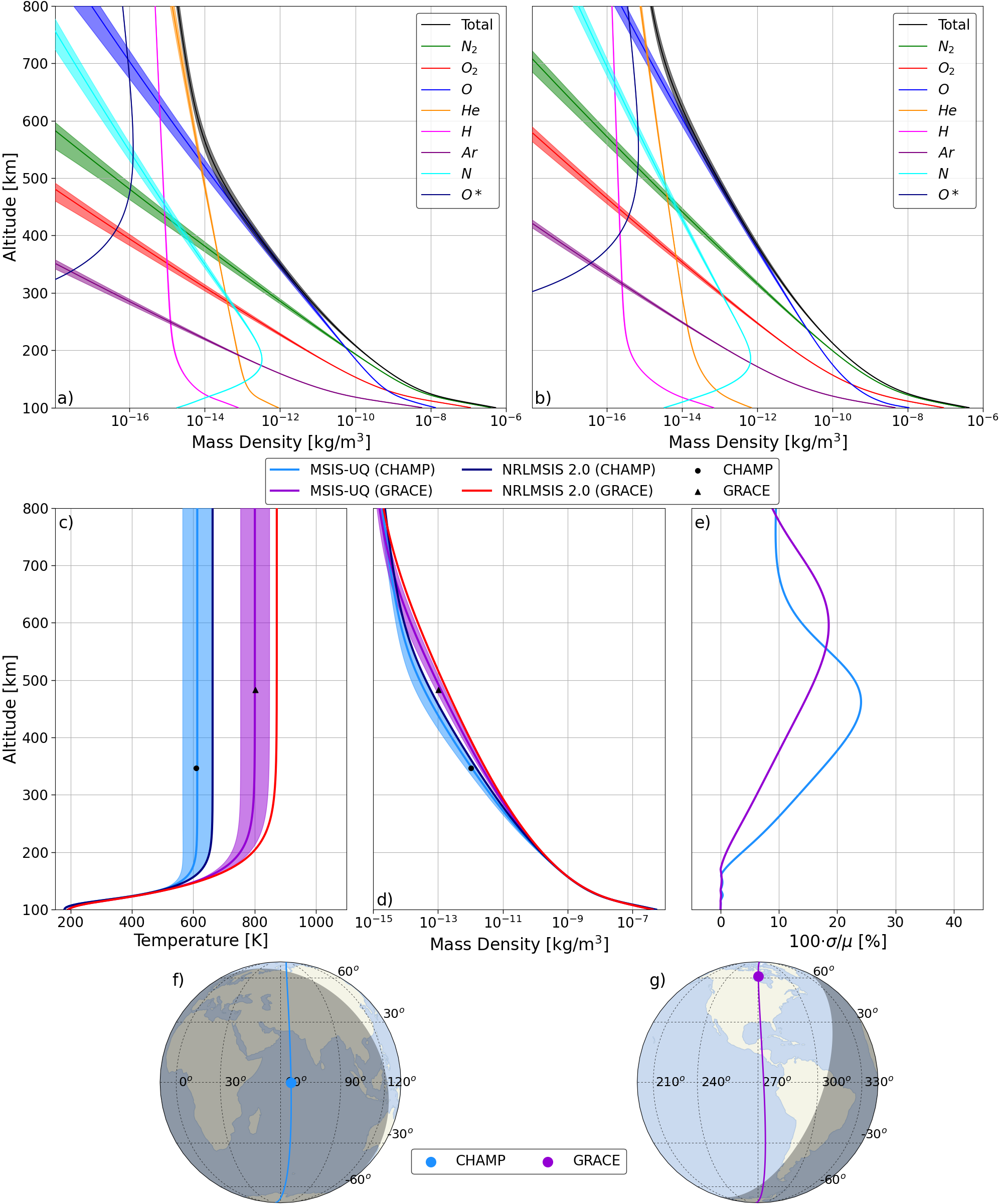}
    \caption{MSIS-UQ species density profiles for CHAMP (a) and GRACE (b) locations with 1-$\sigma$ bounds, temperature profiles with 2-$\sigma$ bounds (c), total mass density profiles with 2-$\sigma$ bounds (d), 1-$\sigma$ uncertainty normalized by the mean prediction (e), and the paths for CHAMP (f) and GRACE (g) with the current location denoted by the markers.  This was conducted for May 13, 2007 at 21:42.50 UT.}
	\label{f:profile}
\end{figure}

Figure \ref{f:profile} contains uncertainty profiles for MSIS-UQ at CHAMP and GRACE locations on May 13, 2007. There are panels for species density, temperature, mass density, relative uncertainty, and satellite position. Please reference the figure caption and Section \ref{sec:UQmethod} for details.

Panels (a) and (b) show the species density profiles at CHAMP and GRACE positions, respectively. The uncertainty bounds provide valuable information on the impact of exospheric temperature uncertainty on the uncertainty of local species. For example, one can investigate the Oxygen (O) to Helium (He) transition for various locations and conditions. Panel (a) shows that at CHAMP's position, this transition is occurring somewhere in the region of 507 to 552 km (1-$\sigma$) while at GRACE's position, the transition may occur between 688 and 738 km (1-$\sigma$). Other insights can be gained such as which species are most impacted by exospheric temperature at a given location/altitude. Note: only 1-$\sigma$ bounds are shown here to prevent artifacts at low-values caused by the semi-logarithmic scale. The scale also causes the bounds to appear to be not-centered about the mean.

Panels (c), (d), and (e) provide information on the local temperature and mass density with uncertainty. In panel (c), MSIS-UQ severely shifts the exospheric temperature prediction and brings it closer to the estimates of CHAMP and GRACE; in both cases NRLMSIS 2.0 overpredicts temperature. The uncertainty in temperature is unobservable below 130 km and grows until it reaches the asymptotic temperature between 250 and 300 km. The uncertainty bounds and mean remain unchanged above these altitudes. Note that the CHAMP and GRACE temperature estimates are for $T_\infty$ but we show them at their current altitude as the temperature has converged. In panel (d), we see different trends in mass density. Again the uncertainty is minimal below approximately 200 km and begins to increase for a few hundred kilometers. The overprediction of temperature in NRLMSIS 2.0 results in higher than observed density by CHAMP and GRACE around 350 and 475 km, respectively. MSIS-UQ provides a more accurate density predictions at the satellite locations. Panel (e) shows the 1-$\sigma$ uncertainty with respect to mean density. This shows different model behavior between the two locations. At CHAMP's location, the uncertainty increases to 24\% around 460 km and decreased until around 700 km where it settles to 9\%. At GRACE's location, the uncertainty continues to increase until it reaches 18\% at 600 km where it begins to decrease.

\subsection{Enhanced Storm and Post-Storm Modeling}\label{sec:cooling}

\begin{figure}[htb!]
	\centering
	\small
	\includegraphics[width=\textwidth]{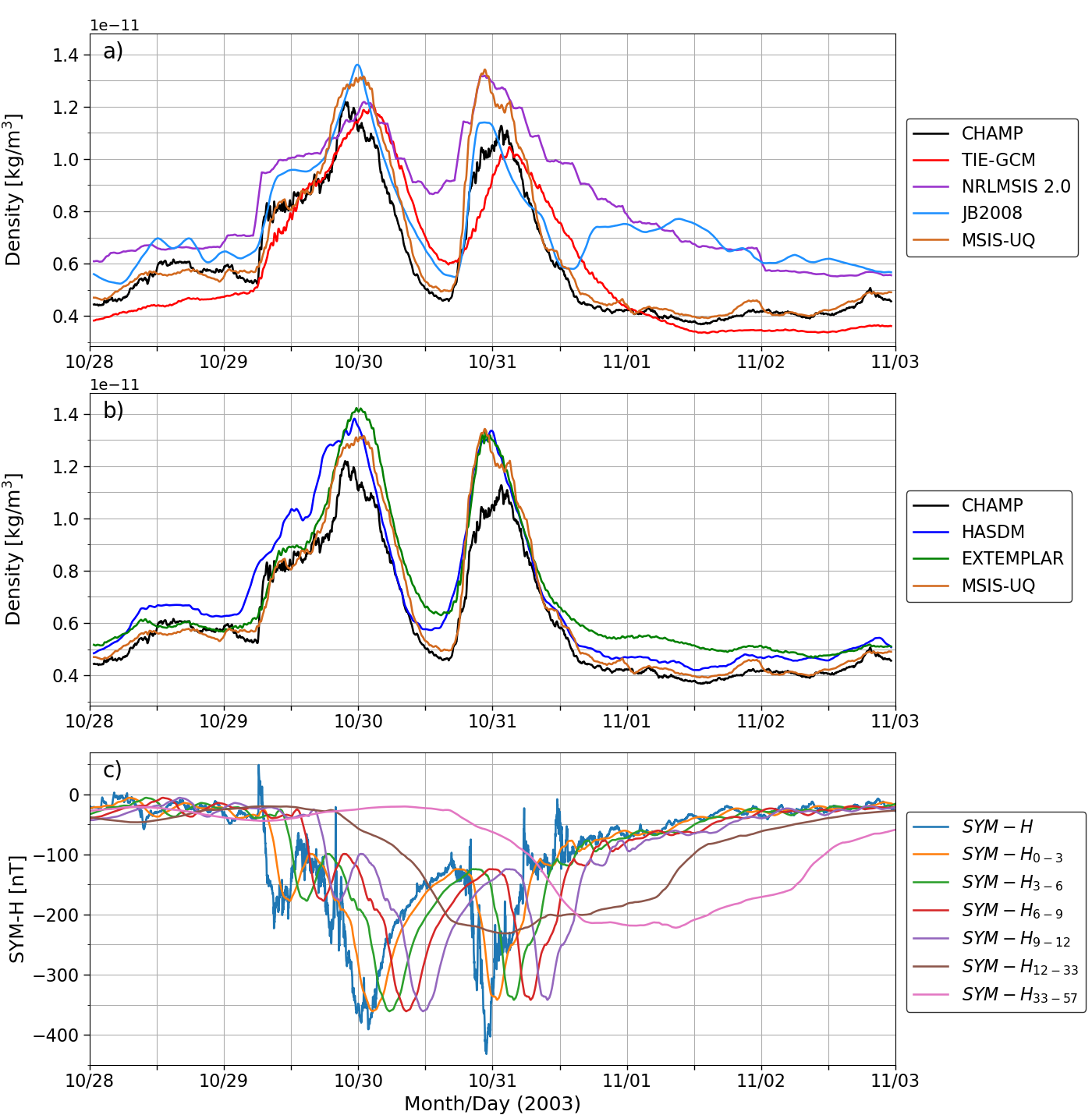}
    \caption{Orbit-average density for TIE-GCM, NRLMSIS 2.0, JB2008, MSIS-UQ, and CHAMP (a), orbit-average density for HASDM, EXTEMPLAR, MSIS-UQ, and CHAMP (b) and the associated \textit{SYM-H} time-series inputs (c).}
	\label{f:storm}
\end{figure}

Geomagnetic storms remain a challenge in modeling the thermosphere. Even further, post-storm characteristics vary between models. To highlight this, we show orbit-average density along CHAMP's orbit during the 2003 Halloween storm for CHAMP, TIE-GCM, NRLMSIS 2.0, JB2008, HASDM, EXTEMPLAR, and MSIS-UQ in Figure \ref{f:storm}. Time history \textit{SYM-H} inputs for MSIS-UQ are displayed in panel (c) for reference.

\begin{figure}[htb!]
	\centering
	\small
	\includegraphics[width=\textwidth]{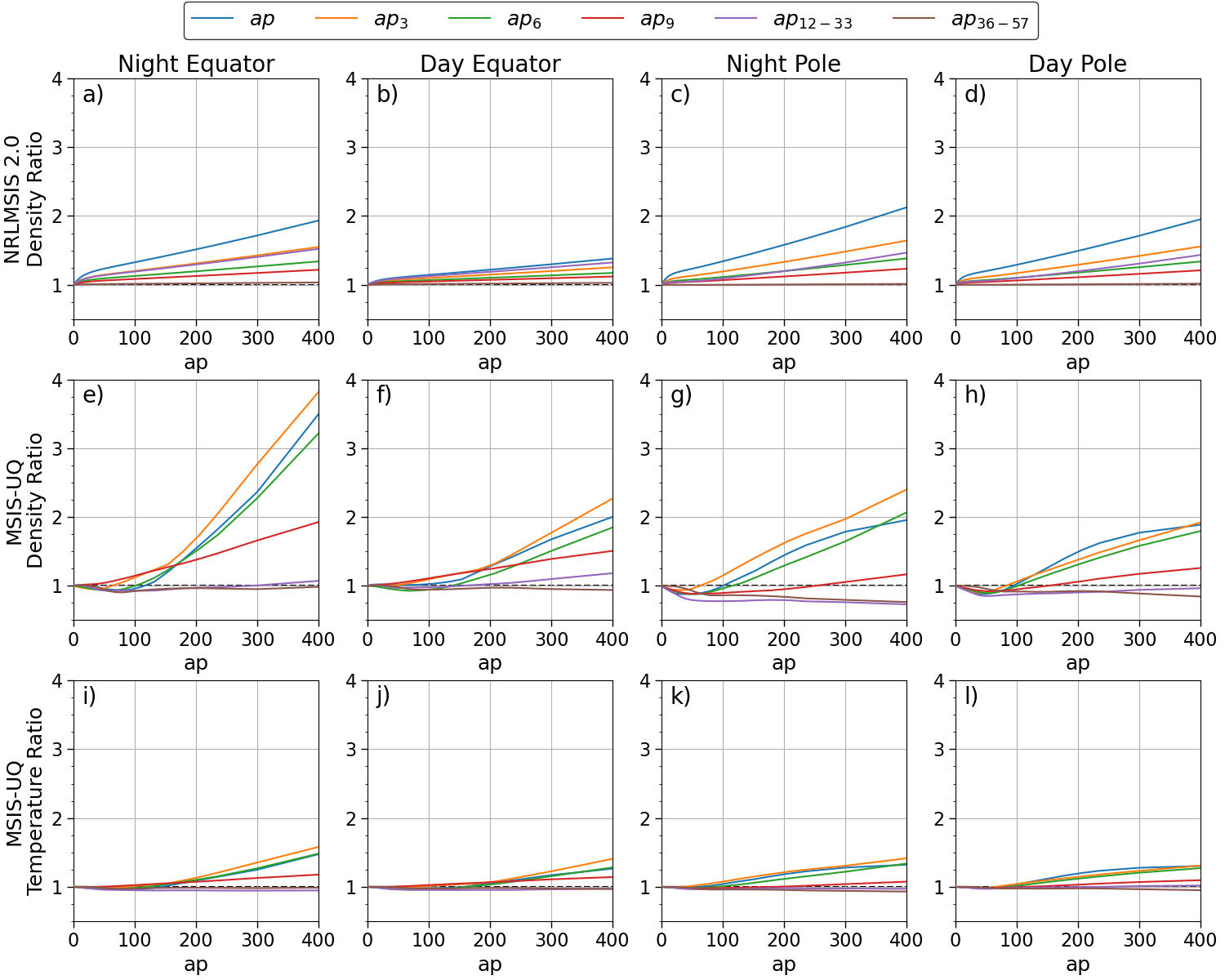}
    \caption{Density ratios for NRLMSIS 2.0 (a--d) and MSIS-UQ (e--h) with the corresponding temperature ratios for MSIS-UQ (i--l).}
	\label{f:lag}
\end{figure}

In the pre-storm period (10/28--10/29), there is significant variability in model outputs. During the first peak of the storm, the models show similar trends but are mostly above the CHAMP density estimates. In the lull between the two peaks (10/30--10/31), all models show density decreases but to varying extents. NRLMSIS 2.0 shows the smallest density decay during this period, but due to the exospheric temperature corrections in MSIS-UQ, the density falls to CHAMP levels. For the second storm, all models overestimate density with the exception of TIE-GCM and JB2008. In the post-storm period (10/31--11/03), NRLMSIS 2.0 does not follow the trend observed by CHAMP -- a sudden and severe decrease in density. JB2008 shows this briefly before overpredicting density to the extent of NRLMSIS 2.0. The MSIS-UQ follows the post-storm overcooling trends observed by CHAMP, showing the impact the exospheric temperature corrections to NRLMSIS 2.0 have. We now conduct the overcooling study from Licata et al. \cite{scienceML} to observe the characteristics in the model without effects from other drivers. These results are shown in Figure \ref{f:lag}, and correspond to Figure 2 in Licata et al. \cite{scienceML}.

Panels (a)--(d) in Figure \ref{f:lag} show the density ratios for NRLMSIS 2.0 while increasing each time-series \textit{ap} value independently. We observe linear relationships between \textit{ap} and density for all of the time-series inputs. The current 3-hourly \textit{ap} value has the strongest impact on density and, of these four locations, never generates a density ratio greater than 2.25. These results show that in NRLMSIS 2.0, the relative importance of each \textit{ap} value becomes less important the further back from epoch it is with \textit{ap\textsubscript{36-57}} having minimal impact on density. The exception to this statement is \textit{ap\textsubscript{12-13}}. At the day equator (panel (b)), it causes the second-most positive density ratio. When \textit{ap\textsubscript{12-13}} is very large, it causes a strong relative increase at all locations considered here. This could explain the overprediction observed in Figure \ref{f:storm}. The second row (panels (e)--(h)) shows the results for MSIS-UQ or the change in NRLMSIS 2.0 when using \textit{SYM-H} time-histories to provide $T_\infty$ correction. The trends shown in these panels contradict many observations when using NRLMSIS 2.0 alone. For example, at 3/4 locations, \textit{ap\textsubscript{3}} causes the largest density ratios, even being nearly twice as large at the night equator. MSIS-UQ also shows a nonlinear relationship between geomagnetic activity and density which is not necessarily seen in NRLMSIS 2.0 alone. 

MSIS-UQ also enforces the idea of negative density ratios -- density decreasing while the geomagnetic drivers are increasing. This is most pronounced at the two pole locations. When the least recent geomagnetic drivers (\textit{ap\textsubscript{12-33}} and \textit{ap\textsubscript{36-57}}) become large, the density becomes up to 25\% lower than when they are set to zero. This overcooling was seen in Figure \ref{f:storm} and is observed in the satellite density data, therefore becoming present in MSIS-UQ. Another interesting trend is seen at low levels on geomagnetic activity particularly in panels (g) and (h). When any of the \textit{ap} values increase from 0 up to 50--100, the density decreases. This seems counter-intuitive but could be caused by the approach of the study. When \textit{ap} is being considered, for example, the \textit{ap} and \textit{SYM-H} values are set to 56 and -50 \textit{nT}, respectively. When \textit{ap} = 0, this would represent the time immediately after moderate geomagnetic activity while \textit{ap} = 56 would represent sustained moderate geomagnetic activity. The model shows that when the conditions abruptly return to quiet values, the density increases -- likely only temporarily. The bottom row (panels (i)--(l)) show the temperature ratios from MSIS-UQ corresponding to the middle row. The general trends are the same between temperature and density; however, the difference comes from the magnitude. The relative changes in temperature result in much stronger changes in density. There are negative temperature ratios, but they are much less prominent due to the consistent scaling.

In SM2, we provide a movie of global density evolution for the 2003 Halloween storm at 400 km. This contains density maps for TIE-GCM, NRLMSIS 2.0, EXTEMPLAR, and MSIS-UQ. Prior to the storm, all models have a similar diurnal structure. This is quickly disrupted at the onset of the storm when both TIE-GCM and MSIS-UQ show strong auroral density enhancements. Throughout the storm, these are the only two models that display significant global disruptions to the structure of the thermosphere at this altitude. The thermosphere expands and contracts both longitudinally and latitudinally for these models. EXTEMPLAR -- a linear model based on the same dataset as MSIS-UQ -- models an abnormal thermosphere, but the movement is more longitudinal \textbf{or} latitudinal with some wave-like oscillations about the equator. During this storm, NRLMSIS 2.0 exhibits general increases in density, but the structure is well-preserved. There are some slow-moving variations although it does not resemble the dynamics exhibited by the physics-based model.

\section{Summary and Conclusions}

In this work, we developed an exospheric temperature model for NRLMSIS 2.0 with uncertainty quantification. When these exospheric temperatures are input to NRLMSIS 2.0, the errors with respect to satellite density estimates are reduced from approximately 45\% to 20\% (Figure \ref{f:TVT}). The relative error distributions for MSIS-UQ have lower variance and bias when compared to both NRLMSIS 2.0 and HASDM. The MSIS-UQ uncertainty estimates proved to be well-calibrated to the satellite density data with a tendency to underestimate the uncertainty bounds.

The uncertainty estimates were closely examined for a given time where CHAMP and GRACE were at unique locations in terms of local time and latitude (Figure \ref{f:profile}). The uncertainty bounds for species densities showed potential for scientific value when considering relative abundances or the uncertainty associated with the O-He transition region. Instead of having a specific altitude where He takes over as the dominant constituent, we observed a 1-$\sigma$ interval of 45-50 km where this may occur, depending on geographical location. Other panels in Figure \ref{f:profile} highlight the improvement in temperature and density prediction with the MSIS-UQ $T_\infty$ predictions. Not only is the biased reduced, the uncertainty estimates can be used to inform decision making (e.g. collision probability estimation). In panel (e), we observed the effect of uncertain exospheric temperature on the relative uncertainty in density as a function of altitude, highlighting the ability to provide different uncertainty ranges as a function of position.

We also investigated the relationship between geomagnetic activity and density. We compared the density of multiple models to CHAMP during the 2003 Halloween storm (Figure \ref{f:storm}). MSIS-UQ portrayed similar characteristics to the satellite estimates, particularly after the storm when there was the most disagreement between the models. This was further explored in the time-lag study in Figure \ref{f:lag}. By individually varying the time-series geomagnetic model drivers, we can observe the models' reaction in density. This showed NRLMSIS 2.0 has a very linear between geomagnetic activity and density. Furthermore, the more recent the driver, the more it impacts density predictions. For MSIS-UQ, we observed nonlinear relationships between many of the time-lagged drivers and density, and the current \textit{ap} value was rarely the most closely tied to density enhancements. We also observed post-storm overcooling effects due to the negative density ratios when the time-lagged drivers were increased. A video showing the evolution of density at 400 km during the 2003 Halloween storm (SM2) provided an example of how MSIS-UQ resembles behavior seen by a physics-based model during extremely nonlinear events, enhancing the capabilities of NRLMSIS 2.0.

MSIS-UQ has value in the community both from a research and operational standpoint. The model can be used to study uncertainty effects on species density, mass density, and temperature on a global scale. It can also be used to investigate global/local mean and uncertainty responses to geomagnetic storms. From an operational perspective, the uncertainty estimates are particularly valuable. MSIS-UQ combines the internal formulation of NRLMSIS 2.0 with calibrated density uncertainties which can be used to obtain more precise collision probability estimates and even uncertainty estimates on satellite re-entry time/location. Further work can be conducted to also tune other temperature profile parameters in NRLMSIS 2.0 to vary the profile in the lower thermosphere.

\section*{Data Statement}

Requests can be submitted for full access to the SET HASDM density database at \url{https://spacewx.com/hasdm/} and all reasonable requests for scientific research are accepted as explained in the rules of road document on the website. The historical space weather indices used in this study can be found at the following links: \textit{F\textsubscript{10.7}}: \url{https://www.spaceweather.gc.ca/forecast-prevision/solar-solaire/solarflux/sx-en.php}, \textit{ap}: \url{https://doi.org/10.5880/Kp.0001}, and \textit{SYM-H}: \url{http://wdc.kugi.kyoto-u.ac.jp/aeasy/index.html}. The remaining solar indices and proxies can be found at \url{https://spacewx.com/jb2008/} in the SOLFSMY.TXT file. Free, one-time only registration is required to access the historical data while nowcasts and forecasts are provided by SET as a data service from data@spacewx.com. The Weimer \cite{PF} Pointing flux data can be accessed at \url{https://doi.org/10.5281/zenodo.3525166}. Programs and files used for ML model development will be made available upon publication. The Supplementary Material (SM) files can be found at \url{https://github.com/rjlicata/msisuqpreprint}.

\section*{Acknowledgements}

PMM gratefully acknowledges support under NSF ANSWERS award \#2149747 (subaward to WVU from Rutgers). All authors acknowledge support by NASA grant \#80NSSC20K1362 to Virginia Tech under the Space Weather Operations 2 Research Program, with subcontracts to WVU and SET. The authors would like to thank Douglas Drob for his insight into the MSIS model. We would like to thank Space Weather Canada for providing and maintaining solar radio emission data, GFZ Potsdam for supplying \textit{ap} archives, and the World Data Center for Geomagnetism in Kyoto for providing \textit{SYM-H} data. The authors would like to acknowledge NASA and DLR for their work in the CHAMP and GRACE missions along with GFZ Potsdam for managing the data.

\bibliographystyle{ieeetr}  
\bibliography{references}

\end{document}